\documentclass [aps, amssymb, twocolumn, showpacs, superscriptaddress]{revtex4-1}
\usepackage{graphicx,graphics, color, epsfig, amsfonts, amssymb, amsmath}
\usepackage{epstopdf}
\usepackage{subfigure}
\usepackage{bm}
\usepackage{times,xspace}
\usepackage[colorlinks,citecolor=blue,linkcolor=red]{hyperref}

\begin{document}

\title{Non-Reciprocal Geometric Wave Diode by Engineering Asymmetric Shapes of Nonlinear Materials}

\author{Nianbei Li$^{1}$, Jie Ren$^{2*}$ \\
\normalsize{$^1$Center for Phononics and Thermal Energy Science, School of Physics Science and Engineering,
Tongji University, 200092 Shanghai, P. R. China}\\
\normalsize{$^2$Theoretical Division, Los Alamos National Laboratory, Los Alamos, NM 87545, USA.}\\
\email{Email: renjie@lanl.gov}}
\date{\today}

\begin{abstract}
{\bf  Unidirectional nonreciprocal transport is at the heart of many fundamental problems and applications in both science and technology. Here we study the novel design of wave diode devices by engineering asymmetric shapes of nonlinear materials to realize the function of non-reciprocal wave propagations. 
We first show analytical results revealing that both nonlinearity and asymmetry are necessary to induce such non-reciprocal (asymmetric) wave propagations.
Detailed numerical simulations are further performed for a more realistic geometric wave diode model with typical asymmetric shape, where good non-reciprocal wave diode effect is demonstrated. 
Finally, we discuss the scalability of geometric wave diodes. 
The results open a flexible way for designing wave diodes efficiently simply through shape engineering of nonlinear materials, which may find broad implications in controlling energy, mass and information transports. }
\end{abstract}

\maketitle

The pursuit of novel devices that are able to manipulate energy,
mass and information transmission has stimulated huge interests
in widespread physical branches such as phononics~\cite{phononics1},
photonics~\cite{photonics1,photonics2,photonics3} and
biophysics~\cite{biophysics1}. As one of the most fundamental and
applicable devices, the wave diode can rectify non-reciprocal wave propagation, in analogy to electronic p-n junctions, where the transmission coefficient is
significantly direction-dependent. As such, the device conducts in the forward direction but insulates in the backward one. There are many interesting
theoretical proposals for thermal diodes~\cite{td1,td2,td3,td4,td5},
optical diodes~\cite{od1,od2}, spin Seebeck diodes~\cite{sse1,sse2} and acoustic diodes~\cite{ad1,ad2}, and
many of them have been successfully verified by
experiments~\cite{ex1,ex2,ex3,ex4,ex5,photonics1,photonics2,photonics3}.

It has been well established that linear structures are not able to break the
reciprocity in reflection-transmission once the time-reversal symmetry is preserved~\cite{rp1}. Therefore, both
nonlinearity and some kinds of symmetry breaking mechanism are
essential for the real non-reciprocal propagation.
Recently,  an interesting design of
wave diode is proposed~\cite{lepri1}, based on the nonreciprocal transmission in a one-dimensional (1D) nonlinear layer structure~\cite{od1,od2}. The model system is
described by the discrete nonlinear Schr\"odinger (DNLS) equation,
which is a reasonable approximation for the wave evolution in
layered photonic or phononic crystals~\cite{dnls1}. The spatially
varying coefficients breaks the mirror reflection symmetry and this
non-homogeneity generates the asymmetric wave propagation together
with nonlinearity. However, the inhomogeneity of coefficients is
not necessary for non-reciprocal wave propagation and also it is usually very hard to design the inhomogeneous coefficient in real materials.

To circumvent the difficulty, in this article we report another novel design by seeking the
symmetry breaking from pure geometric point of view, i.e, engineering asymmetric shapes of homogenous materials.
The advantage of utilizing geometric asymmetry is guaranteed by the
easy and convenient fabrication of functional devices merely through
shape engineering. Different functionality can thus be tunable by
simply tailored adjusting of the geometric shape of the target
device. Therefore, the design of a wave diode device purely induced
by geometric effect is very important in the practical point of view
and would have found potential applications in the future.


To unraveling the underlying mechanism of the nonreciprocal geometric wave diode,
we first analytically solve the transmission problem
for a simple model where the functional nonlinear part consisting of
a triangle structure coupled to a linear ladder lattice in the left
and a linear 1D chain in the right. The derived transmission
coefficients exhibit clear asymmetry depending on the direction of
the incident wave, thanks to the nonlinear triangle structure. The strength and sign of the non-reciprocity can
also be tuned by the intensity of the incident wave.
We then investigate a more realistic system with typical asymmetric
geometry. The system is complicated and analytical result is no
longer available so that we resort to intensive numerical simulations of the propagation of wave packets. The
results clearly demonstrate non-reciprocal wave propagations induced by the asymmetric geometry of nonlinear materials.
We further verify that our proposed
wave diode devices will not change the frequency of the incident wave packet
which would be important in practical applications. Finally, we discuss the scalability of the geometric wave diode.

\begin{figure}
\scalebox{1}[1]{\includegraphics[width=0.7\columnwidth]{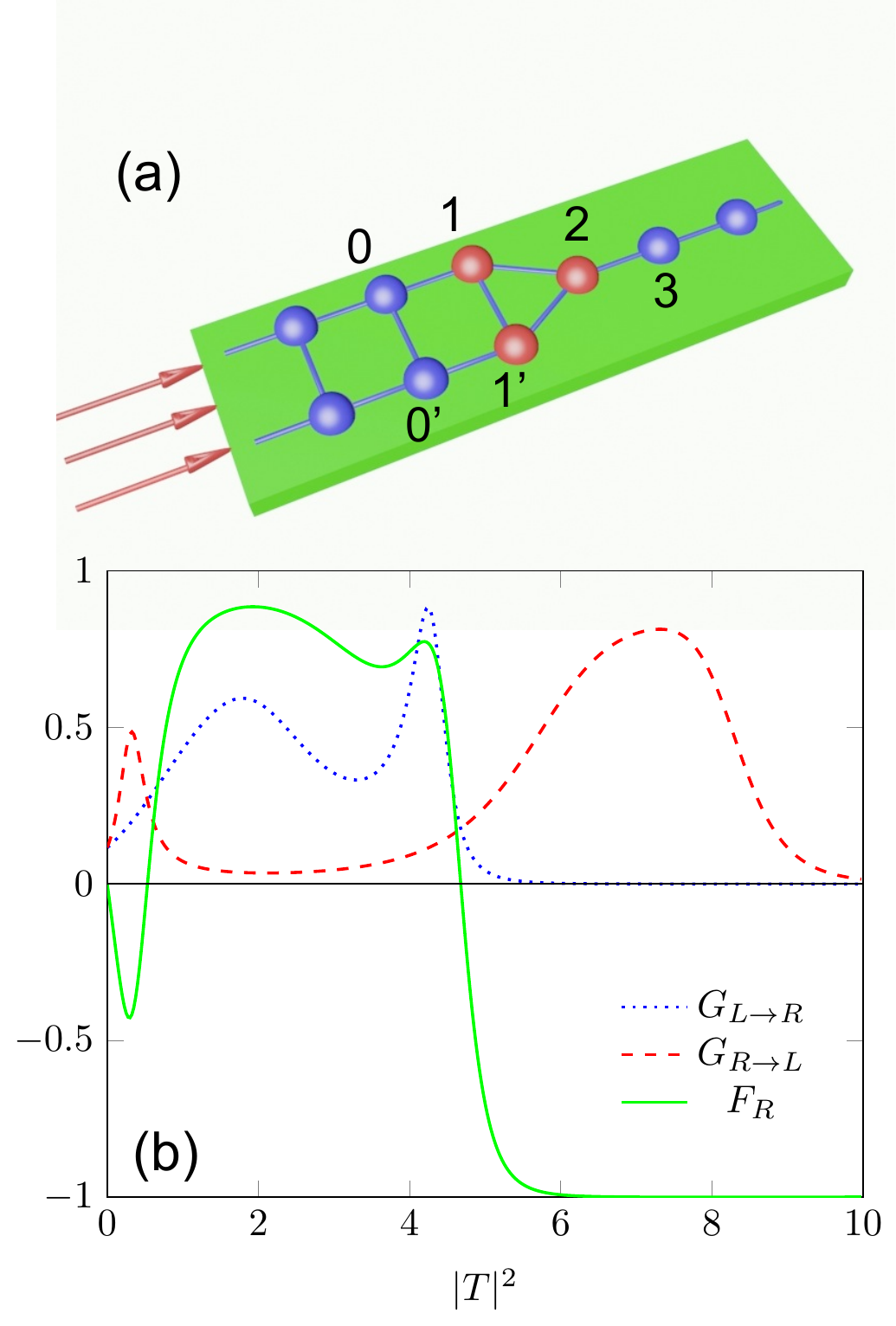}}
\vspace{-5mm}
\caption{(color online) (a) {\bf Sketch of the
geometric wave diode.}  The central $L_c=2$ interfacial layers are
nonlinear, with asymmetric geometry. Its left is coupled with a
semi-infinite linear ladder lattice. The right is connected to a
semi-infinite 1D linear lattice. The linear structures are free of
on-site potentials and the central 2 layer interfaces are with
identical nonlinear strength $g$ and on-site potential $U$.
(b) {\bf Transmission coefficients and rectification factor as
a function of the intensity of transmitted waves.
}
The wave vectors are identical with
$k=1.57$ for both left and right incoming waves. The on-site energy
$U=-3.5$ and nonlinear strength $g=1$ are set for the central two
layers.} \label{fig1}
\end{figure}

\vskip0.25cm\noindent
{\bf Results}\\
{\bf Analytic results for a simple geometric diode.}
Figure \ref{fig1}(a) shows the sketch of a simple geometric wave diode model with
non-reciprocal wave transmission induced by asymmetric shape. The
central triangle lattice with $L_c=2$ interfacial layers is made of
nonlinear materials and plays the role of functionality for the
device. The nonlinear triangle is then coupled with semi-infinite linear
ladder lattice in the left and 1D linear chain in the right,
respectively.

The evolution of wave amplitude $\psi_{l}$ at each site $l$ follows the
time-dependent DNLS equation\cite{dnls}:
\begin{equation}\label{dnls}
i\dot{\psi_l}=(U_l+n_l)\psi_l+g_l|\psi_l|^2\psi_l-\sum_{m_l}\psi_{m_l}
\end{equation}
where $U_l$ and $n_l$ depict the on-site energy and the number of
nearest-neighbors of site $l$, $g_l$ is the nonlinear strength and
$m_l$ denotes the index of nearest-neighbors of site $l$.
The on-site energy and nonlinear strength have homogeneous non-zero
values $U_l=U$ and $g_l=g$ only in the central functional part.
Therefore the symmetry breaking only comes from the geometry.
For the right 1D linear chain, the dispersion relation can be
obtained as $\omega=2-2\cos{k_R}$, where the plane wave solution has
the form of $\psi_l=A_l e^{-i(\omega t-k_R l)}$. While for the
left linear ladder, there are two branches for the energy band.
Depending on whether bilateral amplitudes on the ladder are (anti-)symmetric $\psi_{l'}=\pm\psi_l$ , the dispersion relations
of the plane wave solutions are $\omega=2-2\cos{k_L}$ and
$\omega=4-2\cos{k_L}$, respectively.


We can analytically solve the transmission problem of the wave
propagation by the scattering approach similar as in Ref.
\cite{lepri1}. For the incident wave coming from left to right, the
solutions of the wave amplitudes in the central four layers [see Fig.~\ref{fig1}(a)] can be
expressed with the following boundary conditions:
\begin{eqnarray}\label{bc}
\psi_0&=&\psi_{0'}=I+R, \quad \psi_1=\psi_{1'}=I e^{ik_L} + R e^{-ik_L}\nonumber\\
\psi_2&=&T e^{ik_R}, \quad\quad\quad  \psi_3=\psi_2 e^{ik_R}
\end{eqnarray}
where $I$, $R$ and $T$ denote incident, reflected and transmitted wave
amplitudes, respectively. Without loss of generality, we only
consider the lower dispersion branch of the left ladder lattice, which is
identical to that of the right 1D chain by assuming
$\psi_{l'}=\psi_l$. As such, both left and right linear structures have the same dispersion and the wave
vectors are identical with
$k_{L,R}=k$ under the same energy. Therefore, the forward transmission coefficient $G_{L\rightarrow R}$ from left to right can be derived as the function of transmitted wave intensity $|T|^2$ (see {\bf Methods} for the detailed derivation):
\begin{equation}\label{glr}
G_{L\rightarrow
R}:=\frac{|T|^2\sin{k_R}}{2|I|^2\sin{k_L}}=\frac{8\sin{k_L}\sin{k_R}}{|2-(\alpha-e^{ik_L})(\beta-e^{ik_R})|^2}
\end{equation}
where $\beta\equiv U+3-\omega+g|T|^2$ and
$\alpha\equiv U+2-\omega+g|T|^2(\beta^2+1-2\beta\cos{k_R})/4$. The ratio $\sin k_R/\sin k_L$ appears in the definition as a renormalization factor because the group velocities of incident waves of the left and right leads might not be the same (if the upper energy branch of the left side is chosen, in our case). The factor 2 in the denominator is the consequence that there are two incident channels from the left due to the ladder structure. One can readily verify the conservation condition $2|I|^2\sin k_L=2|R|^2\sin k_L+|T|^2\sin k_R$.
In the same manner, the backward transmission coefficient $G_{R\rightarrow L}$ from right to left can also be obtained as:
\begin{equation}\label{grl}
G_{R\rightarrow
L}:=\frac{2|T'|^2\sin{k_L}}{|I'|^2\sin{k_R}}=\frac{8\sin{k_L}\sin{k_R}}{|2-(\alpha'-e^{ik_L})(\beta'-e^{ik_R})|^2}
\end{equation}
with $\alpha'\equiv U+2-\omega+g|T'|^2$ and
$\beta'\equiv U+3-\omega+g|T'|^2(\alpha'^2+1-2\alpha'\cos{k_R})$.
One can also readily verify the conservation in the backward direction that $|I'|^2\sin k_L=|R'|^2\sin k_L+2|T'|^2\sin k_R$. The factor 2 is due to the fact that there are two transmitted channels to the left.
Note we use prime to denote the quantities in the backward direction.

Although the forward and backward transmission coefficients $G_{L\rightarrow R}$ and $G_{R\rightarrow L}$ have similar expressions, the non-identical behavior between parameters $(\alpha,\beta)$ and $(\alpha',\beta')$ implies the existence of direction-dependent transmission coefficients. For plane waves with same intensities, 
the incident waves will be scattered differently by the central nonlinear triangle due to its asymmetric shape in respect to the forward and backward waves, which leads to the non-reciprocal  propagation.
To quantify the magnitude of the non-reciprocity, a rectification factor $F_R$ can be defined as:
\begin{equation}
F_R=\frac{G_{L\rightarrow R}-G_{R\rightarrow L}}{G_{L\rightarrow R}+G_{R\rightarrow L}}
\label{eq:FR}
\end{equation}
where a non-zero value of $F_R$ emerges for the non-reciprocal wave propagation and reaches to its maximal asymmetry with values $F_R=\pm 1$.

We plot $G_{L\rightarrow R}$ and $G_{R\rightarrow L}$ as well as $F_R$ in Fig. \ref{fig1}(b), as a function of the transmitted wave intensities $|T|^2=2|T'|^2$~\cite{note1} according to Eq. (\ref{glr}) and (\ref{grl}).
It can be seen that the dependence of $G_{L\rightarrow R}$ and $G_{R\rightarrow L}$ on transmitted intensities are almost opposite to each other. As a result, significant non-reciprocal effect of wave propagation emerges and the rectification factor $F_R$ can even change its sign as the modulation of transmitted wave intensities. We would like to remark that analytically we obtain the nonreciprocal transmission because $\alpha\neq\alpha', \beta\neq\beta'$ that are consequences of both the nonlinearity and the asymmetric geometry. If the shape is symmetric, we of course have symmetric transmissions. If the nonlinearity vanishes $g\rightarrow0$, even with keeping the asymmetric geometry, we can see analytically that $\alpha=\alpha'$ and $\beta=\beta'$ so that the reciprocal transmissions are also recovered.

\begin{figure}
\scalebox{1}[1]{\includegraphics[width=0.7\columnwidth]{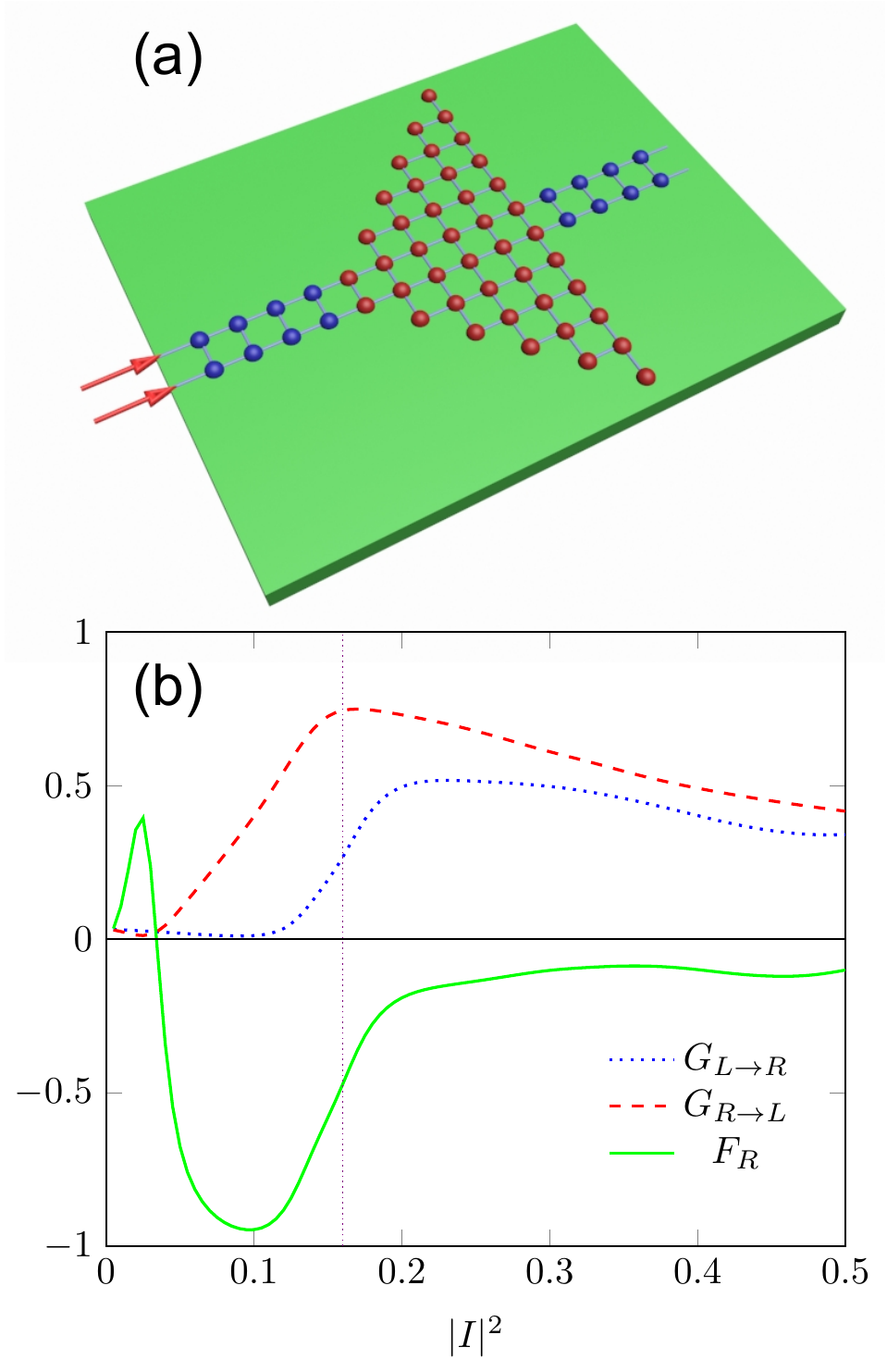}}
\vspace{-5mm}
\caption{(color online). (a) {\bf Sketch of a more
realistic geometric wave diode.} The central $L_c$ layers are nonlinear, with
asymmetric shape. The left and right are coupled with the same
semi-infinite linear ladder lattices. The linear structures are free
of on-site potentials and the central layers are with homogeneous
nonlinear strength $g$ and on-site potential $U$.
(b) {\bf Transmission coefficients and rectification factor as
a function of the intensity of incident waves.}
The initial wave intensities are the same for
both directions with $|I|^2=|I'|^2$. The absolute values of wave vectors are identical
with $|k|=1.57$ for both left and right incoming waves. The central
part has $L_c=6$ layers with homogeneous on-site energy $U=1$ and
nonlinear strength $g=1$. Both of the left and right linear ladders have $L_l=L_r=500$ layers. The width of the initial wave packet is
$d=100$ and the central position of the initial wave packet is $l_0=250$ $(750)$ for left (right) incident wave. The vertical
line denotes the initial wave intensities $|I|^2$ used in Fig.
\ref{fig:evo-m2}.
}
\label{fig:model2}
\end{figure}

\vskip0.25cm\noindent
{\bf Simulation results for a more realistic geometric diode.}
We now consider a more realistic model with typical asymmetric geometry displayed in Fig. \ref{fig:model2}(a). The central functional part with layer index $1\leq l \leq L_c$ is nonlinear in nature and has asymmetric shape. The same left and right linear ladder lattices are coupled with the central part. The system is too complicated to be tackled analytically. Therefore, numerical simulations will be performed to investigate the wave propagation in such a setup.

We consider the propagation of a pulse or say a Gaussian wave packet with the initial condition:
\begin{equation}
\psi_l(t=0)=I e^{-\left(\frac{l-l_0}{d}\right)^2+ik l}
\end{equation}
where $d$ and $l_0$ is the width and central position of the wave packet, respectively. The numerical transmission coefficients $G_{L\rightarrow R}$ and $G_{R\rightarrow L}$ are defined as the sum of transmitted density norms $\sum_{l>L_c}|\psi_l(t)|^2$ at the end of time evolution normalized by the sum of initial density norms $\sum_{l<1}|\psi_l(t=0)|^2$.
In Fig. \ref{fig:model2}(b), we plot the numerical transmission coefficients $G_{L\rightarrow R}$ and $G_{R\rightarrow L}$ as well as the numerical $F_R$ as a function of incident wave intensities $|I|^2$. Although the profiles of forward and backward transmission coefficients exhibit similar behaviors, a lag between them is enough to give rise to a profound non-reciprocal wave diode effect.

\begin{figure}
\scalebox{1}[1]{\includegraphics[width=1\columnwidth]{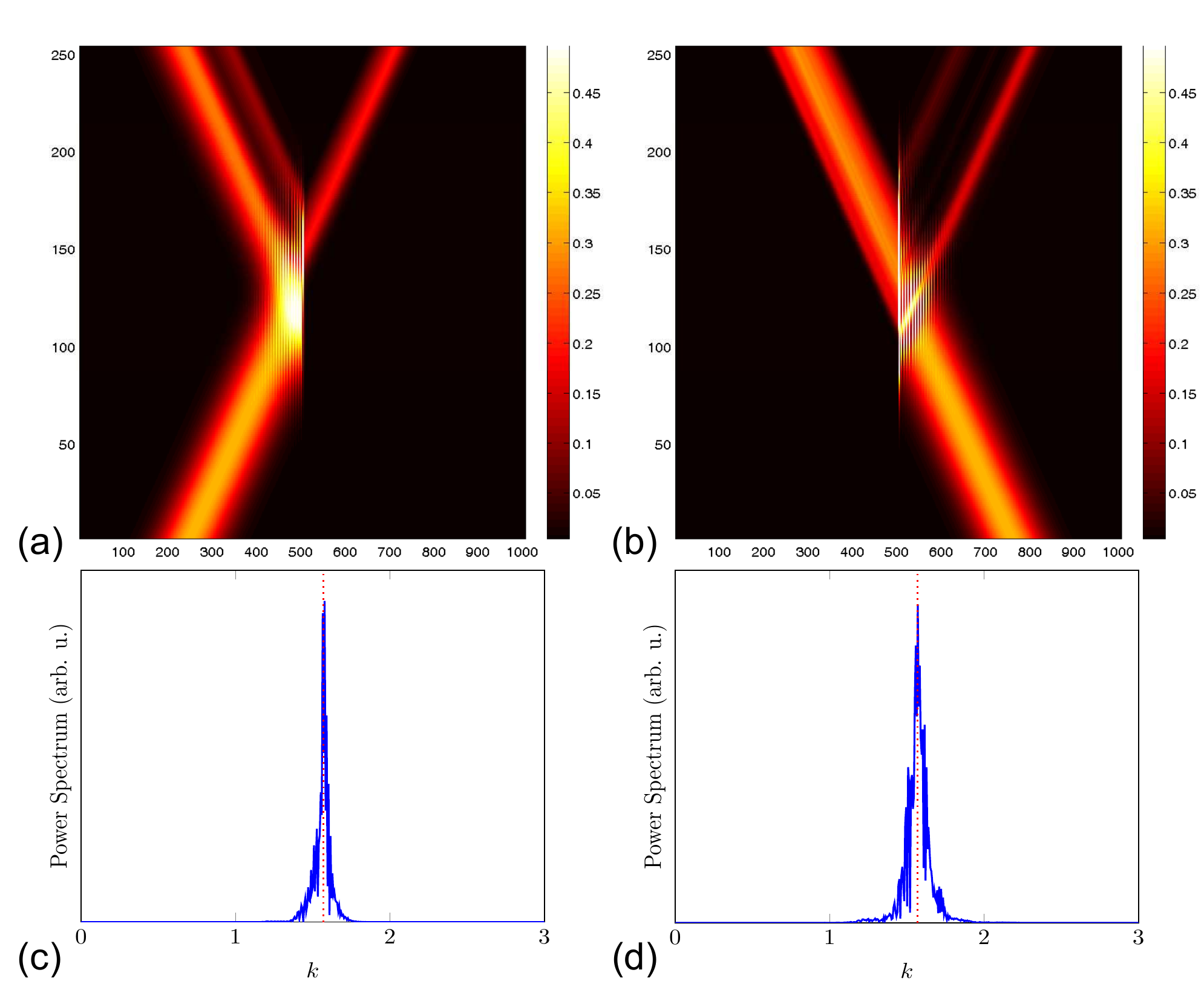}}
\vspace{-7mm}
\caption{(color online) (a,b) {\bf Time evolution of left and right incoming Gaussian wave
packets demonstrates the device as a wave diode.} The parameter setup is the same as in Fig. \ref{fig:model2}
with the initial wave intensities $|I|^2=0.16$ as marked by the
vertical line there.
(c,d) {\bf Power spectra of the real part of $\psi_l$ at the
initial and ending time of the evolution for left and right incoming
wave packets.} The initial wave vectors are
$|k|=1.57$, represented by the vertical dotted lines in left
and right panels. The ending time of the evolution is $t=253$.
}\label{fig:evo-m2}
\end{figure}

Figure \ref{fig:evo-m2}(a,b) displays one typical example of time evolution for the same initial wave packets with opposite incident direction. In the forward direction, most of the incident waves have been reflected by the nonlinear central part. While in the backward direction, most of the incident waves can pass through the nonlinear central part. The interplay between nonlinearity and asymmetric geometry within the central functional part yields the non-reciprocal wave propagations. The asymmetric behavior can be modulated by the incident wave intensities and wave vector $k$. More importantly, the sign and magnitude of this non-reciprocity can also be manipulated by the geometric shape of the central part.


To further check whether the frequency of wave packet can be shifted or not after the scattering, we plot the power spectra of the wave packet at the end of time evolution in Fig. \ref{fig:evo-m2}(c,d). Although nonlinear resonances can be generated by the central part, the reflected and transmitted wave packets almost maintain their original frequencies. This effect is important for many practical applications since usually we do not want the frequency of the original signal to be altered after passing through the functional devices.

\begin{figure}
\scalebox{1}[1]{\includegraphics[width=0.7\columnwidth]{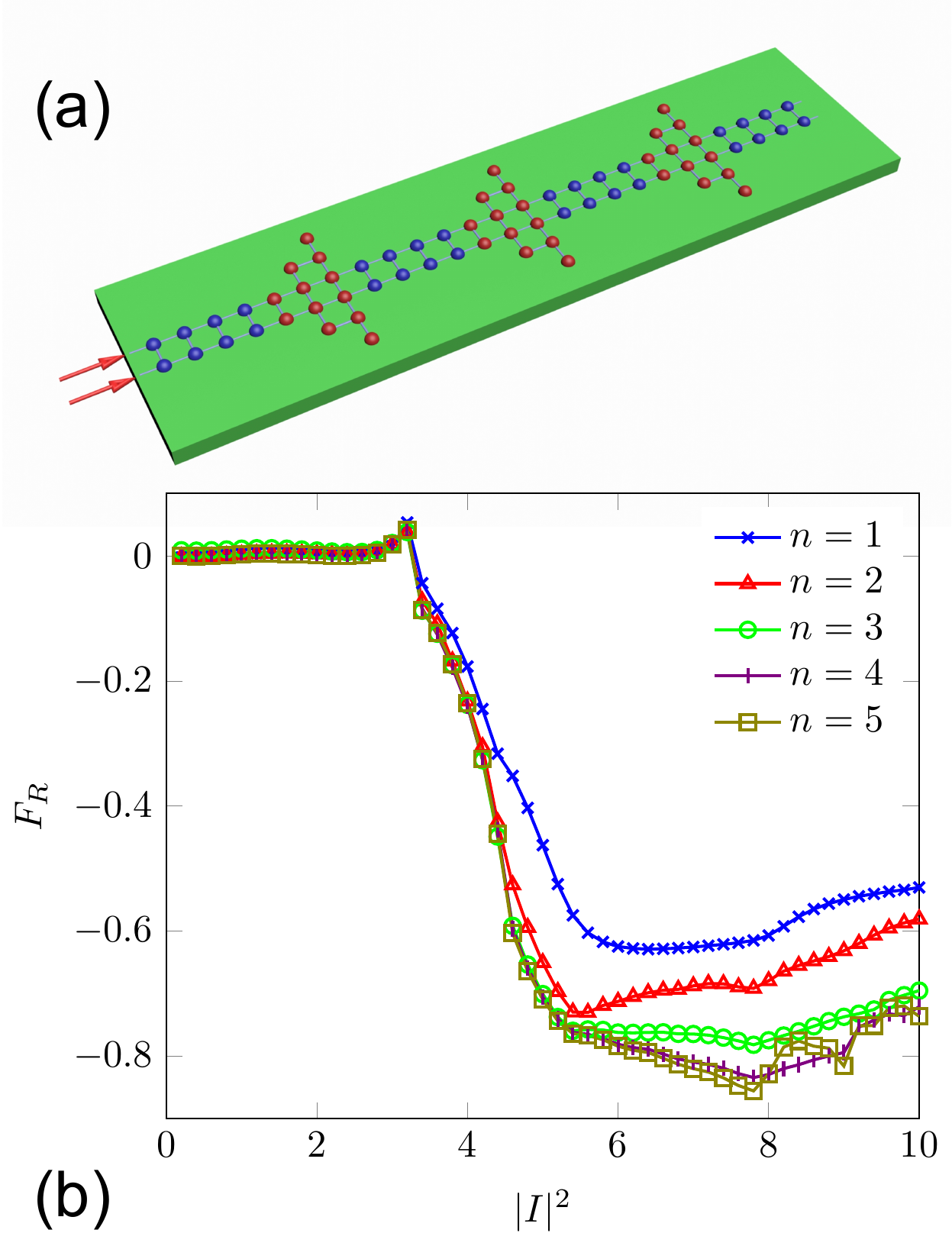}}
\vspace{-4mm}
\caption{(color online) (a) {\bf Sketch of a geometric wave diode linked in series.} The $n$ diodes are linked in series to enhance the rectification factor.
(b) {\bf Rectification factor as
a function of the number of diodes linked in series.} The central part for each diode has $L_c=3$ nonlinear layers with homogeneous on-site energy $U=-1$ and nonlinear strength $g=0.1$. The linear ladder between neighboring diodes has $L_{ll}=50$ layers and the most left and right linear ladders has $L_l=L_r=1000$ layers. The wave vectors and width of initial wave packets are $|k|=0.8$ and $d=30$, respectively. The central positions of the initial wave packets are in a distance of $250$ layers to both ends and the ending time is determined when the travelling distance of the central peak equals to $L-500$ where $L$ is the total number of layers of the whole system.}\label{fig:scale}
\end{figure}

Finally, we discuss the scalability of the rectifying function of geometric wave diodes by linking them in series. As shown in Fig.~\ref{fig:scale}, when scaling up to multiple diodes, the rectification is enhanced. This can be understood by the following very rough but intuitive reasoning: Assuming $n$ geometric diodes in series are independent and not interact with each other, then their forward transmission coefficient is $n$ times multiplication of the singe one, expressed as $G_{L\rightarrow R}^n$. Accordingly, their backward transmission coefficient  will be $G_{R\rightarrow L}^n$. As such, the total rectification factor for the $n$ geometric diodes expresses as $F^n_R=(1-(\frac{G_{R\rightarrow L}}{G_{L\rightarrow R}})^n)/(1+(\frac{G_{R\rightarrow L}}{G_{L\rightarrow R}})^n)$. Readily, it can be proved that increasing $n$ will always enhance $F^n_R$ to its extremum values $\pm1$. However, this is just a rough estimation for ideal cases. In reality, the situation is more complicated: the enhancement of the rectification will saturate at some optimal number ($5$ in Fig.~\ref{fig:scale}), or decrease, or even reverse the sign, depending on the wave vector $k$ and intensities (not shown). More importantly, scaling up the diode devices will steadily suppress the transmission since its value is always smaller than $1$, i.e., $G^n$ decreases with $n$ since $G<1$ generally. Therefore, in practice, one need find the optimal number of geometric wave diodes to balance  demand of both the transmission and the rectification.

\vskip0.25cm\noindent
{\bf Discussion}\\
Geometric phonon and electron diodes have been studied both theoretically and experimentally~\cite{GPD1,GPD2,GPD3,GED1,GED2,GED3}. However,  to our best knowledge so far there is no existing exact results to clearly expose the underlying physics, especially for the more general geometric ``wave'' diode, before this work. In particular, the role of nonlinearity (many-body interaction beyond quadric order) to the geometric diode has not been appreciated.

Therefore, it is worthwhile to emphasize here that the asymmetry alone in linear (quadric) systems cannot induce non-reciprocal wave propagation. Asymmetry must come to play together with nonlinearity to give rise to non-reciprocal wave diodes. In all our studied systems, we have either analytically or numerically checked that the effect of non-reciprocal wave propagation will totally disappear if the nonlinearity is tuned to be zero while keeping all other asymmetric settings. These would explain why the electric and thermal rectifications measured in the asymmetric graphene and its oxide systems are too low~\cite{GED2,GPD3} [both are below 0.13 according to our Eq.~(\ref{eq:FR})], because in these graphene-related systems electrons and phonons are weakly interacting so that the nonlinearity (high order many-body interaction) is nearly absent.

The interplay between some kind of symmetry breaking mechanism and nonlinearity is essential to realizing non-reciprocal wave propagation. Ref. \cite{lepri1} breaks the symmetry by setting spatially varying inhomogenous coefficients to induce asymmetric propagation. Here we proposed a novel design of the so called geometric wave diode by tailoring the asymmetric shape of the central functional part of the device.
The functional nonlinear part of asymmetric shape can be conveniently fabricated in practice and easily be replaced by other shapes with different or even reversed rectifications. Therefore, the most manipulability of the non-reciprocal wave propagation can be solely achieved by geometry engineering of the central functional part.

In conclusion, we have shown the non-reciprocal geometric wave diode by engineering asymmetric shapes of nonlinear materials. For a simple model, analytical results have been obtained indicating that the non-reciprocal wave propagation can be induced by the asymmetric shape in nonlinear materials. Numerical simulations have been performed to study a more realistic diode model with typical geometric asymmetry. Profound non-reciprocal propagations of wave packets have been verified with tailored design of interplay between nonlinearity and geometric asymmetry. Most importantly, the sign and magnitude of the non-reciprocity of the wave propagation can be manipulated by shaping the central functional part into different geometric structures.

Since the DNLS has found widespread implications in optics, cold atoms and spin electronics, the  proposed design in this article might find its applications in controlling the light propagation in nonlinear optical wave guides, cold atom transport in atomtronics~\cite{atom1,atom2}, or spin wave transport in spin caloritronics~\cite{sse2}. In the future, it would be interesting to explore how the asymmetric geometry optimizes the rectification, which may have a deep connection to the quantum chaos~\cite{chaos} with nonlinearity.

\vskip0.25cm\noindent
{\bf Methods}\\
In order to analytically solve the transmission problem for the first model,
we find the traveling wave solution of $\psi_l\sim e^{-i\omega t}$
for Eq. (\ref{dnls}) as:
\begin{equation}\label{dnls-a}
\omega
\psi_l=(U_l+n_l)\psi_l+g_l|\psi_l|^2\psi_l-\sum_{m_l}\psi_{m_l}
\end{equation}
Apply the above equation to site $2$ in Model I, we obtain:
\begin{equation}
\omega\psi_2=(U+3)\psi_2+g|\psi_2|^2\psi_2-(\psi_1+\psi_{1'}+\psi_3)
\end{equation}
it can be regrouped by knowing that $\psi_1=\psi_{l'}$ and
$\psi_3=e^{ik_R}\psi_2$ from the boundary conditions of Eq.
(\ref{bc}):
\begin{equation}
\psi_1=\frac{1}{2}(\beta-e^{ik_R})\psi_2
\end{equation}
with
\begin{equation}
\beta=U+3-\omega+g|T|^2
\end{equation}
Apply Eq. (\ref{dnls-a}) to site $1$, we obtain:
\begin{equation}
\omega\psi_1=(U+3)\psi_1+g|\psi_1|^2\psi_1-(\psi_0+\psi_{1'}+\psi_2)
\end{equation}
After regrouping and substitution we can obtain:
\begin{equation}
\psi_0=\left(\frac{1}{2}\alpha(\beta-e^{ik_R})-1\right)\psi_2
\end{equation}
with
\begin{equation}
\alpha=U+2-\omega+g|T|^2(\beta^2+1-2\beta\cos{k_R})/4
\end{equation}
From the first two boundary conditions in Eq. (\ref{bc}), the
incident amplitude $I$ can be expressed as:
\begin{equation}
I=\frac{e^{-ik_L}\psi_0-\psi_1}{e^{-ik_L}-e^{ik_L}}
\end{equation}
Substitute the expression of $\psi_0$ and $\psi_1$ into above
formula and noticing that $\psi_2=Te^{ik_R}$ in Eq. (\ref{bc}), we
can obtain the relation between incident and transmitted amplitude:
\begin{equation}
I=e^{ik_R}\frac{e^{-ik_L}\left(\alpha(\beta-e^{ik_R})/2-1\right)-(\beta-e^{ik_R})/2}{e^{-ik_L}-e^{ik_L}}T
\end{equation}
The transmission coefficient of Eq. (\ref{glr}) can thus be derived
analytically. For the reversed direction, the transmission
coefficient of Eq. (\ref{grl}) can also be derived with the same
consideration.

For the numerical simulation, we apply the symplectic PQ method
introduced in the appendix of Ref. \cite{Josh} to integrate the time
dependent DNLS equations. It is a $\mbox{SBAB}_2$ based symplectic
algorithm\cite{sbab2}. The dimensionless time step $\Delta t=0.02$
has been used in all the numerical simulations.

 \vskip0.25cm\noindent
{\bf Acknowledgements}\\
N.L. acknowledges the supports from Shanghai Supercomputer Center, the National Natural
Science Foundation of China, Grant No. 11205114, and Shanghai Rising-Star Program with Grant No. 13QA1403600. J.R. thanks the support by the National Nuclear Security Administration of the U.S. DOE at LANL under Contract No. DE-AC52-06NA25396 through the LDRD Program.

%

\clearpage
\newpage


\begin{thebibliography}{10}
\bibitem{note1} Here, we choose the same transmitted intensities for the convenience, because the transmission is analytically expressed in terms of transmitted intensities. We have checked at the stable regime that the same incident amplitudes also produce the nonreciprocal transmission. In fact, in the following simulations, we choose transmissions at the same incident amplitudes for the comparison.
\end{thebibliography}
\end{document}